\documentclass[prd,superscriptaddress,amsfonts,amssymb,twocolumn,amsmath,showpacs]{revtex4-2}
\usepackage{bm}
\usepackage{amsfonts}
\usepackage{tensor}

\usepackage{latexsym}
\usepackage[utf8]{inputenc}
\usepackage{graphicx}
\usepackage{amsmath}
\usepackage{palatino}
\usepackage{mathpazo}
\usepackage[british]{babel}
\usepackage{hhline}
\usepackage{multirow}
\usepackage{textcomp}
\usepackage{float}
\usepackage{booktabs}
\usepackage{dcolumn}
\usepackage{lipsum} 
\usepackage{mdframed}
\usepackage[]{mdframed}
\usepackage{hhline}
\usepackage{multirow}
\usepackage{ragged2e}
\usepackage{hyperref}
\hypersetup{colorlinks,citecolor=blue}
\hypersetup{colorlinks=true,linkcolor=red,filecolor=magenta,    urlcolor=cyan}
\usepackage{amsmath}
\usepackage{xcolor}
\usepackage{orcidlink}
\usepackage{epsfig}
\usepackage{caption}
\usepackage{subcaption}
\usepackage{commath}

\def\jnl@style{\it}
\def\aaref@jnl#1{{\jnl@style#1}}

\def\aaref@jnl#1{{\jnl@style#1}}

\def\aj{\aaref@jnl{AJ}}                   
\def\apj{\aaref@jnl{ApJ}}                 
\def\apjl{\aaref@jnl{ApJ}}                
\def\apjs{\aaref@jnl{ApJS}}               
\def\apss{\aaref@jnl{Ap\&SS}}             
\def\aap{\aaref@jnl{A\&A}}                
\def\aapr{\aaref@jnl{A\&A~Rev.}}          
\def\aaps{\aaref@jnl{A\&AS}}              
\def\mnras{\aaref@jnl{Mon.~Not.~Roy.~Astron.~Soc.}}             
\def\prd{\aaref@jnl{Phys.~Rev.~D}}        
\def\prc{\aaref@jnl{Phys.~Rev.~C}}  
\def\prl{\aaref@jnl{Phys.~Rev.~Lett.}}    
\def\qjras{\aaref@jnl{QJRAS}}             
\def\skytel{\aaref@jnl{S\&T}}             
\def\ssr{\aaref@jnl{Space~Sci.~Rev.}}     
\def\zap{\aaref@jnl{ZAp}}                 
\def\nat{\aaref@jnl{Nature}}              
\def\aplett{\aaref@jnl{Astrophys.~Lett.}} 
\def\apspr{\aaref@jnl{Astrophys.~Space~Phys.~Res.}} 
\def\physrep{\aaref@jnl{Phys.~Rep.}}      
\def\physscr{\aaref@jnl{Phys.~Scr}}       
\def\commat{\aaref@jnl{Comm.~Math.~Phys.}}              
\def\science{\aaref@jnl{Science}}               
\def\cqg{\aaref@jnl{Classical Quant.~Grav.}}            
\def\jpcs{\aaref@jnl{JPCS}}                                     
\def\ijmpd{\aaref@jnl{Int.~J.~Mod.~Phys.~D}}                    
\def\grg{\aaref@jnl{Gen.~Relat.~Gravit.}}               
\def\rpp{\aaref@jnl{Rep.~Prog.~Phys.}}          
\def\npa{\aaref@jnl{Nucl.~Phys.~A}}        
\def\lrr{\aaref@jnl{Living Rev.~Rel.}}                   
\def\jcap{\aaref@jnl{J.~Cosmology Astropart.~Phys.}}    
\def\rmp{\aaref@jnl{Rev.~Mod.~Phys.}}   
\def\epjc{\aaref@jnl{Eur.~Phys.~J.~C}} 
\def\plb{\aaref@jnl{~Phy.~Lett.~B}} 
\def\mpla{\aaref@jnl{Mod.~Phy.~Lett.~A}} 
\def\arxiv{\aaref@jnl{arxiv.org}}

\allowdisplaybreaks[1]

\begin{document}

\title{Bouncing Cosmology in Interacting Scalar–Torsion Gravity}

\author{S. A. Kadam\orcidlink{0000-0002-2799-7870}}
\email{sak.scos@jspmuni.ac.in;
\\k.siddheshwar47@gmail.com}
\affiliation{Department of Mathematics, School of Computational Sciences, JSPM University Pune-412207, India}

\author{A. S. Agrawal\orcidlink{0000-0003-4976-8769}}
\email{asagrawal.sbas@jspmuni.ac.in}
\affiliation{Department of Mathematics, School of Computational Sciences, JSPM University Pune-412207, India}
\begin{abstract}
\textbf{Abstract:}
In this study we demonstrate the interacting teleparallel gravity models, to describe the matter bounce scenario. We discussed two interacting models and find both are suitable choice to describe the bouncing phenomena.
The co-moving Hubble radius, is demonstrated to check the establishment of the matter bounce scenario. All the energy conditions and the behaviour of EoS parameter is analysed. The violation of NEC at bounce epoch is one of the crucial result to establish bouncing behaviour is found to be obeyed. The other energy conditions behaviour is in agreement with the EoS parameter which lies in the phantom region at the bounce epoch in both the models.
\end{abstract}

\maketitle

\section{Introduction}\label{Introduction}
In-depth examination of the standard model of cosmology is currently crucial for comprehending the historical development of the universe's cosmological evolution \cite{weinberg2008,Carroll1992}. Although this model continues to struggle with the initial singularity problem, the inflation does not seem to resolve the issue of the initial singularity \cite{A_Guth, LINDE1982389}. In fact, the analysis seen in Ref.\cite{Borde_1994} indicates that nearly all parts of the inflation region will have a singularity at some point in the history. To address this problem, it is necessary to present a scientific model for the known universe that depicts it as oscillatory. This suggests that our Universe is a consequence of the collapse of a preceding universe \cite{Ashtekar_2006}. This novel concept, known as the bouncing Universe, can be suggested to address the non-singularity of big bang cosmology \cite{Ilyas_2021,Peter_2002}. The interpretation of the bouncing Universe indicates that when the bounce occurs, the Universe transitions into the big bang era, shifting from an initial contracting phase to an expanding phase, wherein the Hubble parameter transitions from $H$ taking values less than $0$ to $H$ taking values grater than $0$, and at the point of the bounce, $H = 0$. One more important tool to test the occurrence of the bouncing cosmology is the existence of matter fields that violate the null energy condition and are widely analysed in the literature Refs. \cite{AGRAWAL2021100863, agrawal2022bouncing, agrawal2022role, agrawal2023bouncing, agrawal2023matter}. A promising approach to addressing the issue of cosmological singularities can be addressed using matter bounce, non-singular bounce, symmetric bounce, super-bounce and other discussed bouncing cosmological solutions \cite{Cai_2011,Bamba_2014,Amani_2016,dela_2018,caruana2020,MISHRA2024138968}. The reconstruction of the models using the bouncing scenario have been employed in the majorly studied modified gravity formalisms namely the modified GR, teleparallel and the symmetric teleparallel gravity (TG) studied in \cite{2022EPJP,Gadbail_2023,Azhar_25,2022EPJP}.

  An alternative equivalant formalism to GR is the teleparallel equivalent of general relativity (TEGR), which features a Lagrangian that is equal to that of TEGR, minus a boundary term \cite{Maluf:2013gaa,Pereira:2013qza}, as a result, it generates identical dynamical equations to GR. In GR to describe gravitation, the Levi-Civita connection is considered , which is curvature-free and torsion-less, meets the non-metricity requirement. In TG framework, the standard metric Levi-Civita connection is substituted with the Weitzenb\"{o}ock connection \cite{Ferraro:2008ey,Weitzenbock1923}, which is characterized by torsion and has zero curvature while adhering to the metricity condition. The metric tensor $g_{\mu\nu}$, serves as the core dynamical entity in GR and several of its extensions. In TG, this tensor can be derived from the tetrad, \( e^{a}_{\mu} \), instead of the metric tetrad serves as the fundamental variable of the theory \cite{Aldrovandi:2013wha}. To review TG formalism refer to \cite{bahamonde:2021teleparallel}. The modifications to the TEGR formalism gains a significant attention to study different observational tensions in the current epoch of the Universe evolution \cite{Di_Valentino_2025}. One of the initial modification to TEGR is $f(T)$ gravity \cite{Ferraro:2008ey,2011JCAP,DUCHANIYA2024f(T)}, where $f(T)$ is the general function of torsion scalar $T$. The $f(T)$ gravity introduces novelty through its second-order field equations, but this benefit comes at the cost that the theory does not maintain Lorentz invariance, as the torsion scalar $T$ is also not invariant under such transformations; additionally, the Ricci scalar and the torsion scalar only differ by a total derivative term \cite{Bahamonde:2015zma}. The $f(T)$ gravity theory gains significant attention due to its capacity to address the cosmic tensions like $H_0, \sigma_{8}$ \cite{Wang_2020,Emmanuel_H}. Moreover $f(T)$ gravity explains the current and early phases of the Universe evolution \cite{Bengochea:2008gz}, solar system test \cite{Iorio:2012cm,Farrugia:2016xcw}, violation of black hole thermodynamics \cite{Miao_2011} and the exact solutions through Noether symmetry \cite{Basilakos:2013rua}.  The particular type of bouncing solutions like super-bounce, loop quantum cosmological ekpyrotic is analysed in \cite{ODINTSOV_2015} where well known gravity models like $f(R), f(T), f(G)$ have been discussed. The scalar tensor perturbations and the matter bounce phenomena are investigated \cite{Cai_2011}. From the available literature on bouncing cosmology, the major work done is in the modified GR formalism, we can find in TG and it's modifications \cite{caruana2020,dela_2018,Azhar_25}, and very few in teleparallel modified gravity formalism containing scalar field \cite{MISHRA2024138968}. This work concentrates on interacting scalar field models with the goal of investigating the matter bounce scenario in the context of a teleparallel interacting scalar field. The matter bounce scenario while considering various potential functions using the scalar tensor $f(T, \phi)$ formalism is studied in \cite{MISHRA2024138968}. Our objective in this work is to analyze the matter bounce scenario with a well-established exponential potential \cite{Copeland_1998} within two specific interacting scalar field models. These models have been reconstructed and studied within the dynamical system framework as outlined in Ref. \cite{Ashmita_2024,Roy:2023uhc}. This work represents one of the earliest attempts to study bouncing cosmology using these models. The action formulation utilized in this research is detailed in \cite{G.Otalora2013JCAPDSA,WEI2012430,Xu_2012,Geng:2011}. 
  
    The work is presented in the sequence which start with the detail representation of the TG formalism in Sec. \ref{TG_Formalism}. The concept of co-moving radius is presented in the support of the bouncing solutions in Sec. \ref{comoving}. The bouncing cosmology for considered two interacting scalar tensor models is presented, in details in subsections \ref{Model_I} and \ref{Model_II}, the EoS parameters have been demonstrated. The energy conditions to validate the bouncing solutions are discussed in detail in Sec \ref{ec}. Finally the results are summerised in the summary and conclusion section in Sec \ref{conclusion}.
\section{INTERACTING Teleparallel Gravity Formalism}\label{TG_Formalism}
The action formula we have discussed here is as presented below \cite{G.Otalora2013JCAPDSA,WEI2012430,Xu_2012,Geng:2011} ,
\begin{equation}
S = \int d^4x\, e \left[ \frac{T}{2\kappa^2}
+ \frac{1}{2}\partial_\mu \phi \, \partial^\mu \phi
- V(\phi) + \xi f(\phi) T \right] + S_m,
\label{eq:action}
\end{equation}
here, $\partial_{\mu} \phi \partial^{\mu} \phi/2$ is the quadratic kinetic term, $S_{m}$ represents the matter component. $\xi$ is the coupling constant with $f(\phi)$ is the general function of canonical scalar field. The variation of the above action \eqref{eq:action} with respect to tetrads presented in \eqref{tetrad} will produce the general form of field equation as follow,
\begin{widetext}
\begin{align}
\Bigg[ e^{-1} e^{a}{}_{\alpha} \partial_\sigma 
\big( e \, e_{a}{}^{\tau} S_{\tau}{}^{\rho\sigma} \big) 
+ T^{\tau}{}_{\nu\alpha} S_{\tau}{}^{\rho\nu}
+ \frac{T}{4}\delta^{~~\rho}_{\alpha} \Bigg]  
\, 2\left( \frac{1}{\kappa^2} + 2\xi f(\phi) \right) 
+ 4\xi S^{~~\rho\sigma}_{\alpha} f_{,\phi} \, \partial_\sigma \phi \notag \\
+ \left( \tfrac{1}{2}\partial_\mu \phi \, \partial^\mu \phi - V(\phi) \right) \delta^{~~\rho}_{\alpha}
- \partial_{\alpha} \phi \, \partial^{\rho} \phi
= \Theta^{~~\rho}_{\alpha}.
\label{eq:field_eq}
\end{align}
\end{widetext}
the term $S^{~~\rho\sigma}_{\tau}$ involved in this equation is the super-potential, can be calculated using,
\begin{equation}\label{4}
    S_{\tau}^{~~\rho \sigma}=\frac{1}{2}(K^{\rho \sigma}_{~~~\tau}+\delta^{\rho}_{\tau}T^{\theta \sigma}_{~~~\theta}-\delta^{\sigma}_{\tau}T^{\theta \rho}_{~~~\theta})\,.
\end{equation}
and the formula for torsion tensor can be expressed as \cite{Farrugia:2018gyz},
\begin{equation}\label{2}
T^{a}_{~~\mu \nu} = \partial_{\mu} e^{a}_{~~\nu}-\partial_{\nu}e^{a}_{~~ \mu} - \omega^{a}_{~~ b\nu}e^{b}_{~~\mu}+\omega^{a}_{~~ b\mu}e^{b}_{~~\nu}.
\end{equation}
here the $\omega^{a}_{~~ b\nu}$ is the spin connection, it's the role is discussed in the detail in the covariant study \cite{Krssak:2015oua}. The torsion $T$ can be calculated using following formulas,
\begin{equation}
T = S_{a}^{\, \, \, \, \, \mu \nu} T^{a}_{\, \, \, \, \, \mu \nu }=\frac{1}{4}\, T^{\rho\mu\nu} T_{\rho\mu\nu}
+ \frac{1}{2}\, T^{\rho\mu\nu} T_{\nu\mu\rho}
- T_{~~\rho\mu}{}^{\rho}\, T^{\nu\mu}{}_{~~\nu},\label{torsioscalar}
\end{equation}
In TEGR framework, the key variable is represented by the tetrad \( e^a_{\mu} \), where the Greek indices denote spacetime indices and the Latin indices indicate tangent space indices,
\begin{align}
    e^{a}_{~~\mu} = diag(1,a(t), a(t), a(t)) \label{tetrad}
\end{align}
where $a(t)$ is scale factor, expresses the evolution of the Universe and is function of cosmic time $t$. The orthogonality condition satisfied by tetrad field can be described as, 
$e^{~~\mu}_ae^b_{~~\mu}=\delta_a^b$. The metric \( g_{\mu\nu} \) can be related to the tetrad in the following way, which establishes a connection between the metric tensor \( g_{\mu \nu} \) and the Minkowski tangent space metric \( \eta_{ab}=\text{diag}(-1,1,1,1) \) through the relation \cite{bahamonde:2021teleparallel},
\begin{equation}
g_{\mu \nu}=\eta_{ab} e^{a}_{\, \, \, \mu} e^{b}_{\, \, \, \nu}. 
\end{equation}
The Ricci Scalar is represented by $R$, while $B={2}e^{-1}\partial_{\mu}(eT^{\alpha\mu})$ denotes a boundary term associated with the divergence of the torsion tensor. Because $B$ is a total derivative, it does not affect the field equations; therefore, the action of TG is entirely equivalent to the Einstein-Hilbert action, the relation of $R$ to TEGR torsion scalar $T$, can be presented as \cite{Bahamonde:2021gfp,Bahamonde:2015zma},
\begin{equation}\label{_R+B}
T = -R + B\,.
\end{equation}
In this study, we examine a flat Friedmann-Lema\^{i}tre-Robertson-Walker (FLRW) background geometry characterized by its metric presented as,
\begin{equation}
ds^{2}=-dt^{2}+a^{2}(t)[dx^2+dy^2+dz^2],   
\end{equation}
with this setup the torsion scalar in \eqref{torsioscalar} can be calculated as, $T=6H^2$, the action formula presented in Eq. \eqref{eq:action} generates the field equations presented as \cite{G.Otalora2013JCAPDSA},
\begin{align}
\rho_{\phi} &= - 6 \, \xi H^{2} f(\phi)+ \tfrac{1}{2}\dot{\phi}^{2} + V(\phi), \label{rho_hi} \\[6pt]
P_{\phi} &= 4 \xi H f_{,\phi}\dot{\phi} 
+ 6 \xi \left( 3H^{2} + 2\dot{H} \right) f(\phi)+ \tfrac{1}{2}\dot{\phi}^{2} - V(\phi) , \label{p_hi} 
\end{align}
Here $\rho_{\phi}, p_{\phi}$ are the energy density and pressure of the scalar field, $H=\frac{\dot{a}(t)}{a(t)}$ is the Hubble parameter. The dot represents the derivative with respect to the cosmic time $t$. The potential we have considered here is exponential potential $V(\phi)=\gamma e^{-\lambda \phi}$ \cite{Copeland_1998,roy2018dynamical} and the scalar field model $f(\phi)=\alpha \dot{\phi}^2$ \cite{G.Otalora2013JCAPDSA}. The variation with respect to scalar field will generate the Klein-Gordon equation as,
\begin{align}
  -\sigma=3H\dot{\phi} + V_{,\phi}  +  6 \xi f_{,\phi} H^{2}+\ddot{\phi}, \label{klein_Gordon} 
\end{align}
The scalar charge $(\sigma)$ represents the interaction between teleparallel dark energy and dark matter, defined by the relationship $\delta \mathcal{S}_m = -e\sigma$ \cite{Gumjudpai_2005} and is taking the form $\sigma= \frac{Q}{\dot{\phi}}$ \cite{G.Otalora2013JCAPDSA}. By reformulating \eqref{klein_Gordon} in terms of and $p_{\phi}, \rho_{\phi}$, we derive the continuity equation for the field as,
\begin{align}
-Q&=\dot{\rho}_{\phi} + 3H \rho_{\phi}(1 + \omega_{\phi}),\\[6pt]
Q&=\dot{\rho}_{m} + 3H \rho_{m}(1 + \omega_{m}). \label{conservation_eq}
\end{align}
Here $\rho_{m}, p_{m}$, represents energy density, pressure for matter including normal matter, baryon and dark matter. The parameter $Q$ represents the interaction between dark matter and dark energy \cite{Ashmita_2024}. In this study we choose the matter bounce scale factor $a(t)=[1+\frac{3 \eta t^2}{4}]^{\frac{1}{3}}$ to analyse the rate of evolution of the  Universe. The Hubble parameter in this case will take the form $H(t)=\frac{2 \eta t}{3 \eta t^2 +4}$ \cite{Lohakare_2022,agrawal2023matter}. This form is well motivated and plays a important role in describing the matter bounce scenario , analysing the phantom dark energy models \cite{Haro_2012}, to explore the loop cosmology in modified TG models like $f(T)$ \cite{de_Haro_2015}. We demonstrate our analysis considering this form in both the models.   

\begin{figure}[H]
    \centering
    \includegraphics[width=0.45\textwidth]{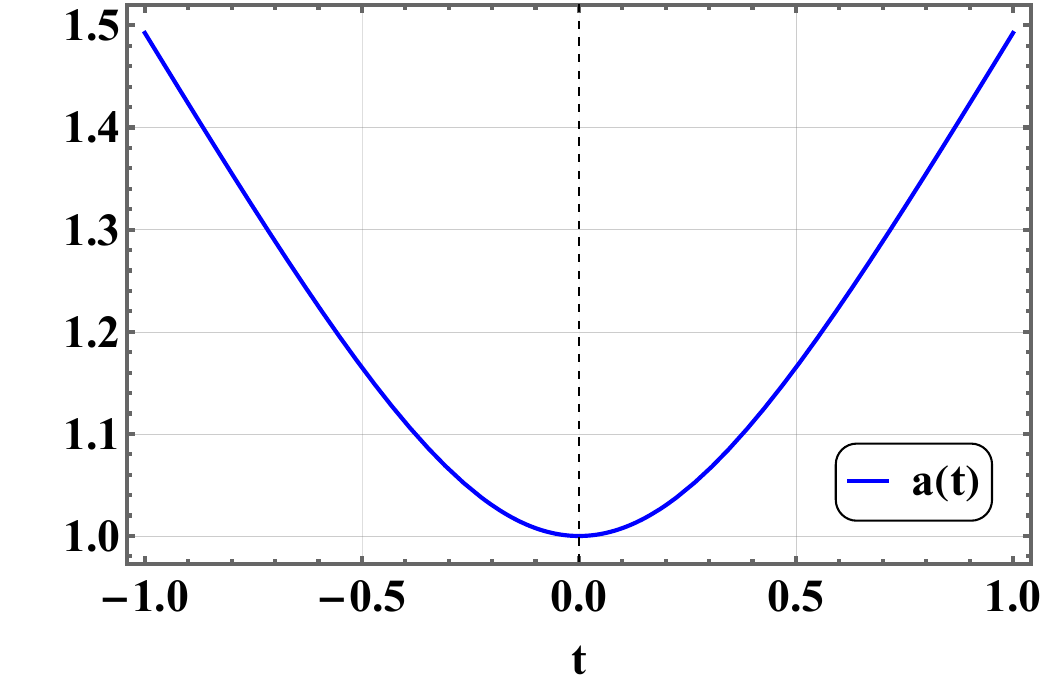} \hfill
    \includegraphics[width=0.45\textwidth]{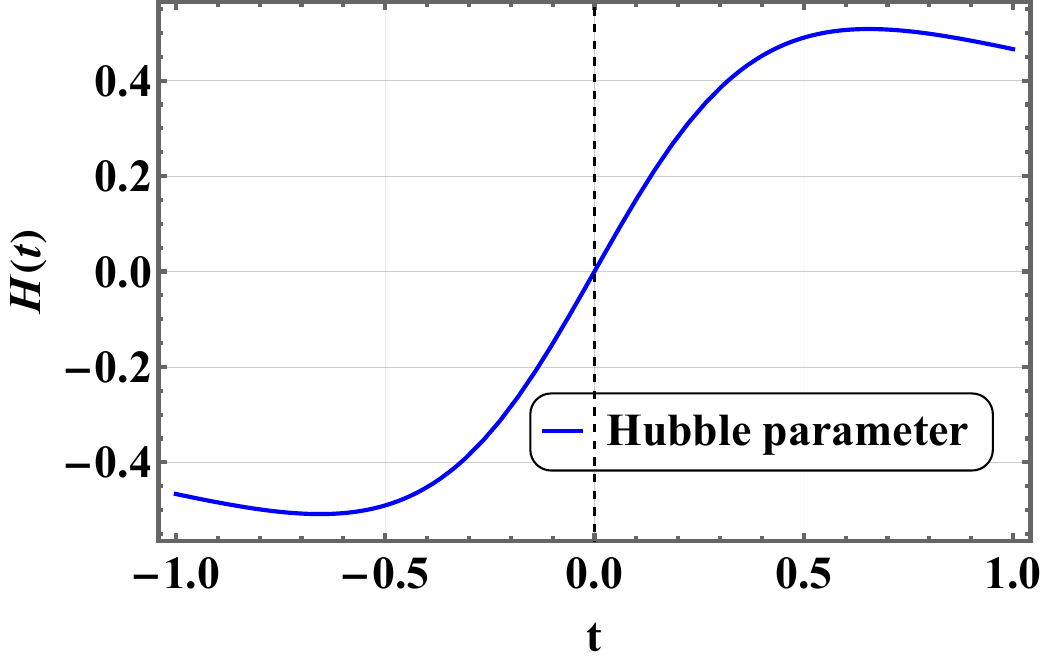}
    \caption{Evolution of the scale factor $a(t)$ and the Hubble parameter $H(t)$ as functions of cosmic time $t$.}
    \label{fig:scale-hubble}
\end{figure}

The figures presented in Fig. \ref{fig:scale-hubble}, illustrate the evolution of the scale factor, which measures the expansion and contraction behavior of the universe. The scale factor demonstrates a contracting behavior in the negative time zone, transitions from contracting to expanding at its minimum value (at \(t=0\)), and then begins to expand in the positive time zone. In contrast, the Hubble parameter, which measures the rate of change of contraction or expansion of the universe, initially shows a negative value, crosses zero at \(t=0\), and becomes positive in the positive time scale and same can be observed in Fig. \ref{fig:scale-hubble}. This behavior of the scale factor $a(t)$, and the Hubble parameter $H(t)$, both are the functions of cosmic time $t$  aligns perfectly with the matter bounce scenario of the universe. This scenario is an alternative to the Big Bang cosmology, suggesting that the universe began with a bounce rather than a Big Bang. Prior to the bounce, the universe was in a contracting phase, and after the bounce, it entered the expansion phase we observe today. Presently, it is thought that the universe experiencing an expansion in an accelerated way. The figures represent both the contracting and expanding phases of the universe, along with this bouncing nature, which also resolves the initial singularity issue that poses a limitation for the Big Bang model.  
\section{Co-moving Hubble Radius}\label{comoving}

\begin{figure}[H]
    \centering
        \centering
        \includegraphics[width=0.4\textwidth]{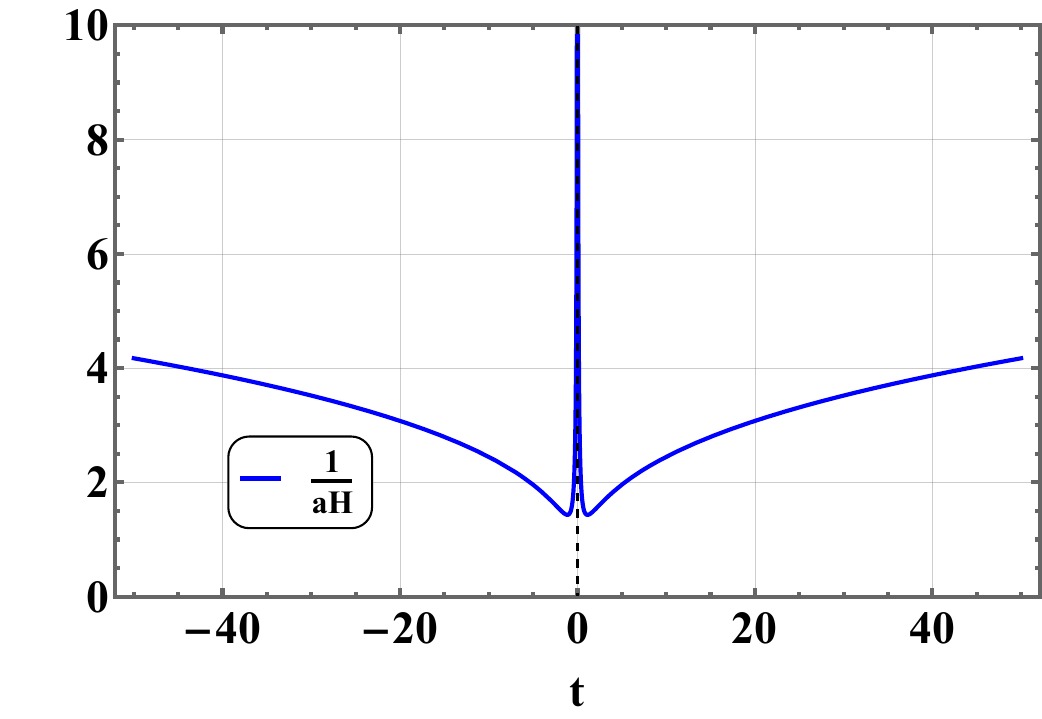}
        \caption{Co-moving Hubble radius}
    \label{fig:CHR}
\end{figure}

In Fig. \ref{fig:CHR} we present the plot of co-moving Hubble radius. The bouncing cosmological models, the Hubble parameter vanishes at the bounce, causing the co-moving Hubble radius $r_h = 1/(aH)$ to diverge. The late-time behavior of the universe accelerating or decelerating can be inferred from the asymptotic evolution of this radius. For certain choices of scale factors, the Hubble radius decreases continuously on both sides of the bounce and approaches zero at late times, indicating an accelerating universe. In contrast, for other scale factor choices, the Hubble radius diverges at late times, signifying a decelerating universe \cite{Odintsov-2020-959}. 

In the following subsections \ref{Model_I}, \ref{Model_II}, we have analysed the models which describes interacting couplings to the dark energy sectors. The study of such models have been obtained previously available in the literature analyse the asymptotic behaviour of warm inflation, dynamics of dark energy and dark matter interaction \cite{Chen_2009,Ashmita_2024,B_hmer_2008,Mimoso_2006,Roy_2019,Roy:2023uhc,Kumar_2019}.

\subsection{$Q=\beta H \dot{\phi}^2$}\label{Model_I}
In this study, $Q$ represents the non-homogeneous part of the Klein-Gordon equation, which is considered under a power-law model of the scalar field.   This first model of interaction was analysed previously in Refs. \cite{Roy_2019,Roy:2023uhc,Ashmita_2024}. The model parameter $\beta$ and $\lambda$ from the exponential potential quantify the strength of the interaction between dark matter and dark energy. In this case, the Klein-Gordon equation presented in Eq. \ref{klein_Gordon}, will take the form,
\begin{eqnarray}
    & &\frac{2 \eta  t \left((\beta +3) \left(3 \eta  t^2+4\right) \dot{\phi}+24 \alpha  \eta  \xi  t \phi\right)}{\left(3 \eta  t^2+4\right)^2}\notag\\
    &+&\gamma  \lambda  \left(-e^{-\lambda  \phi}\right)+\ddot{\phi}=0
\end{eqnarray}

\begin{figure}[H]
    \centering
    \begin{subfigure}[t]{0.4\textwidth}
        \centering
        \includegraphics[width=\textwidth]{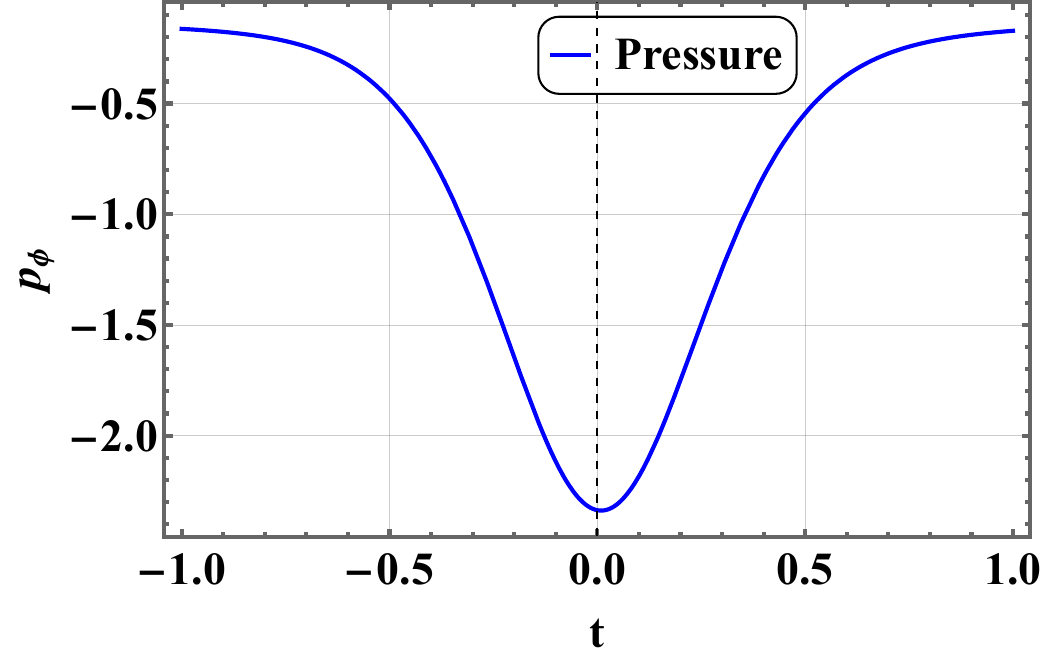}
        \caption{Pressure $p_{\phi}$}
        \label{fig:pressure:m1}
    \end{subfigure}
    \hfill
    \begin{subfigure}[t]{0.4\textwidth}
        \centering
        \includegraphics[width=\textwidth]{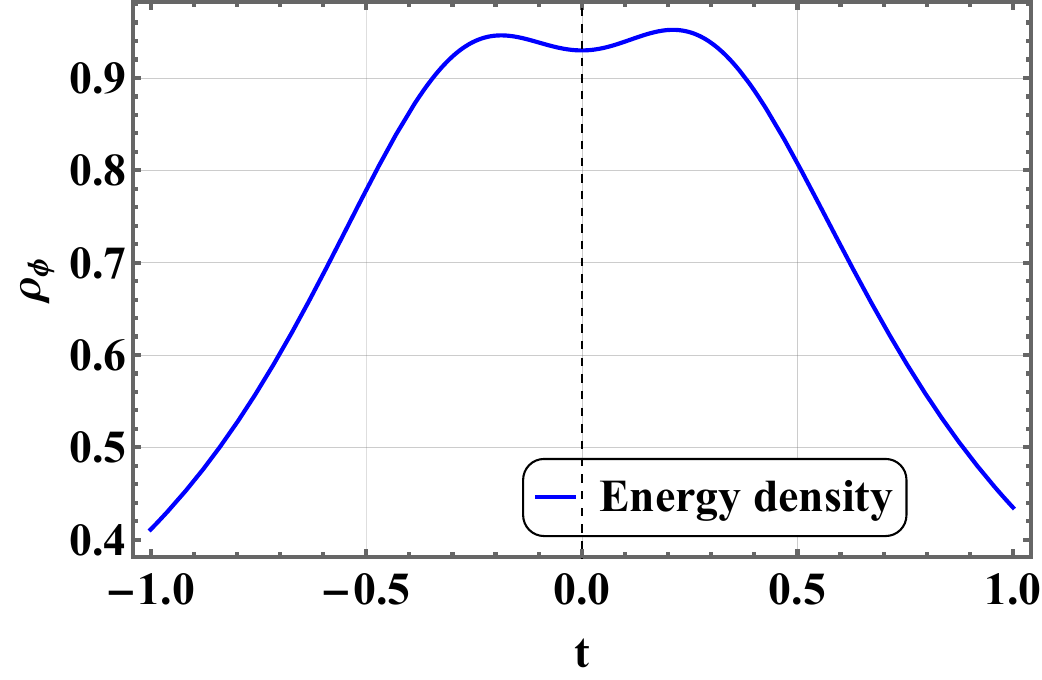}
        \caption{Density $\rho_{\phi}$}
        \label{fig:density:m1}
    \end{subfigure}
    \hfill
    \begin{subfigure}[t]{0.4\textwidth}
        \centering
        \includegraphics[width=\textwidth]{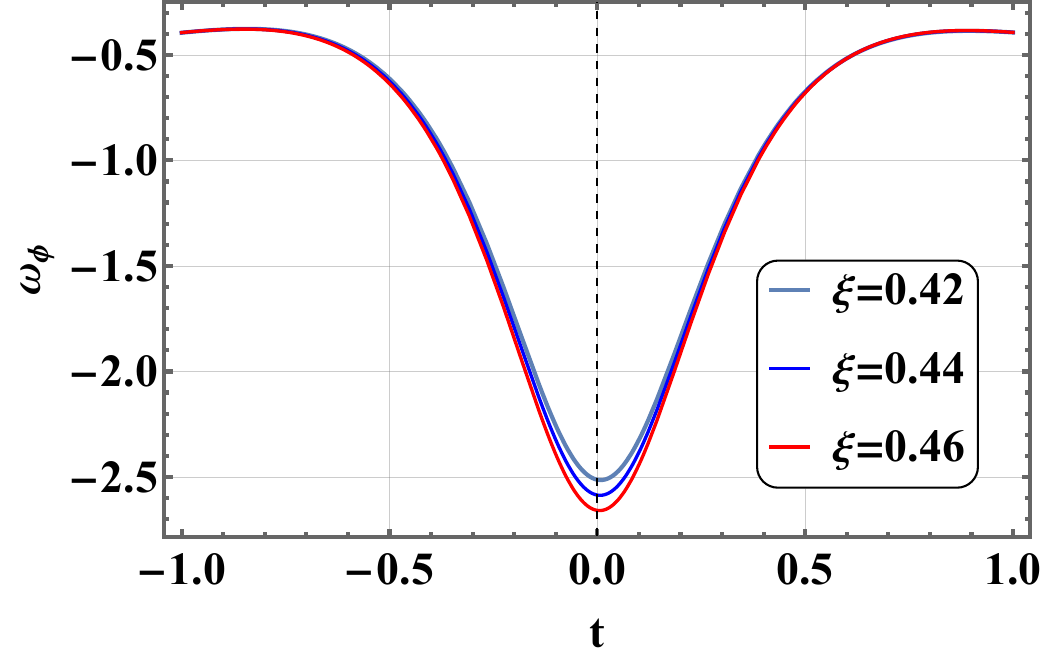}
        \caption{EoS $\omega_{\phi}$}
        \label{fig:eos:m1}
    \end{subfigure}
    \caption{Behavior of pressure, density, and EoS parameter with cosmic time \( t \).}
    \label{fig:prho-eos:m1}
\end{figure}
To investigate the bouncing behavior, we present the plots for pressure, energy density, and the equation of state (EoS) parameter in Fig. \ref{fig:prho-eos:m1}. These dynamic parameters are illustrated with the following parameter values: $\eta = 3.1, \alpha = -0.44, \gamma = 0.2, \beta = 1.3, \lambda = -2.4,$ and $\xi = 0.42$. We obtain the numerical solution for the $\phi$ from the Klein-Gordon equation in each of these cases. 

The pressure corresponding to this configuration, for specific parameter values, is observed to remain negative throughout the cosmic evolution from early to late times. Such negative pressure behavior indicates the presence of an antigravity effect in the Universe. The energy density, on the other hand, exhibits positive behavior. At $t = 0$, a distinct bump is observed; before and after this bump, the density attains its maximum value. As the density increases, the thermal energy also rises, and due to uncontrollable entropy, particles tend to move apart from each other, resulting in the expanding nature of the Universe. The EoS parameter displays a phantom-like behavior in the vicinity of the bounce point. As the Universe evolves away from the bounce epoch, the EoS parameter crosses the $\Lambda$CDM line and subsequently enters the quintessence regime. The EoS parameter has been analyzed for three different values of the model parameter, and it is found that with increasing parameter values, the bump at the bounce epoch decreases while the depth of the well increases. However, within the parameter range $\xi \in (0.42, 0.46)$, the EoS parameter demonstrates the most suitable behavior consistent with the requirements of a bouncing cosmological scenario \cite{AGRAWAL2021100863, agrawal2022bouncing, agrawal2022role, agrawal2023bouncing, agrawal2023matter}.

\subsection{$Q=\tau H \rho_{\phi} \dot{\phi}^2$}\label{Model_II}
The second model is investigated to analyse the dark sector interactions using the observational data studies \cite{Roy:2023uhc,Kumar_2019}. In this section, we showcase the bouncing characteristics of this coupling and examine whether this coupling is the most suitable option for investigating bouncing solutions. In this case the Klein-Gordon equation in Eq. \eqref{klein_Gordon} will take the form,
\begin{eqnarray}
    &-&\frac{24 \alpha  \eta ^2 \xi  t^2 \Phi  (\tau  \phi -2)}{\left(3 \eta  t^2+4\right)^2}+\frac{6 \eta  t \dot{\phi}}{3 \eta  t^2+4}+\gamma  (\tau -\lambda ) e^{-\lambda  \phi}\notag\\
    &+&\frac{1}{2} \tau  \phi^2+\ddot{\phi}=0
\end{eqnarray}

\begin{figure}[H]
    \centering
    \begin{subfigure}[t]{0.4\textwidth}
        \centering
        \includegraphics[width=\textwidth]{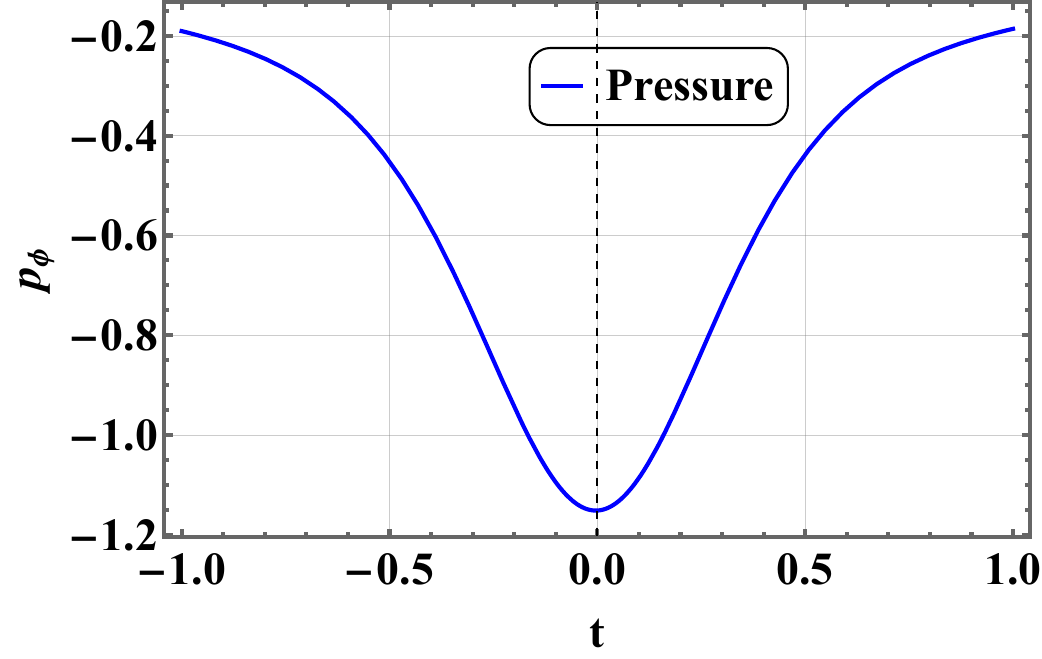}
        \caption{Pressure $p_{\phi}$}
        \label{fig:pressure:m2}
    \end{subfigure}
    \hfill
    \begin{subfigure}[t]{0.4\textwidth}
        \centering
        \includegraphics[width=\textwidth]{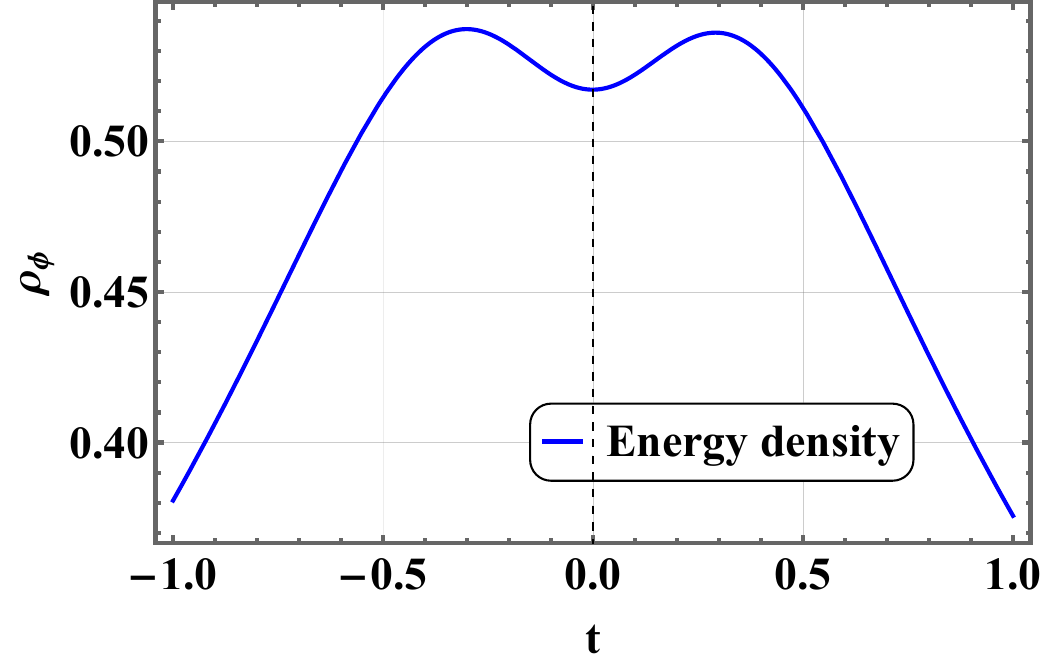}
        \caption{Density $\rho_{\phi}$}
        \label{fig:density:m2}
    \end{subfigure}
    \hfill
    \begin{subfigure}[t]{0.4\textwidth}
        \centering
        \includegraphics[width=\textwidth]{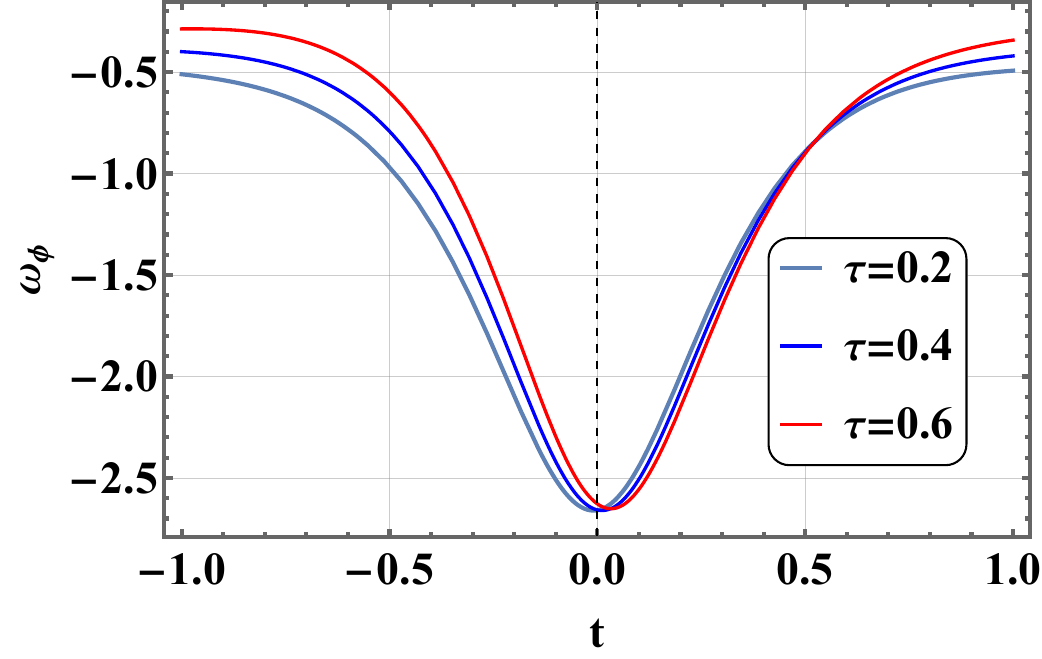}
        \caption{EoS $\omega_{\phi}$}
        \label{fig:eos:m2}
    \end{subfigure}
    \caption{Behavior of pressure, density, and EoS parameter with cosmic time $t$.}
    \label{fig:prho-eos:m2}
\end{figure}
Building on the previous section where we analyzed the behavior of the pressure, energy density, and EoS parameter for the interaction term \(Q = \beta H \dot{\phi}^{2}\), we now consider the alternative coupling \(Q = \tau H \rho_{\phi} \dot{\phi}^{2}
\). Our motivation is to verify whether the small bump observed earlier can be smoothed out within this modified framework. We find that the bump persists in both interaction models and therefore cannot be eliminated by this change. Furthermore, while the first case exhibits an approximately symmetric evolution around the bounce, the present model does not; its behavior is distinctly asymmetric can be analysed through Fig. \ref{fig:prho-eos:m2}. The plots of behaviour of scalar field is presented in the following Fig. \ref{fig:phi:models}, it has been observed that scalar field lies in the positive region for both of interacting models.

\begin{figure}[H]
    \centering
    \includegraphics[width=0.45\textwidth]{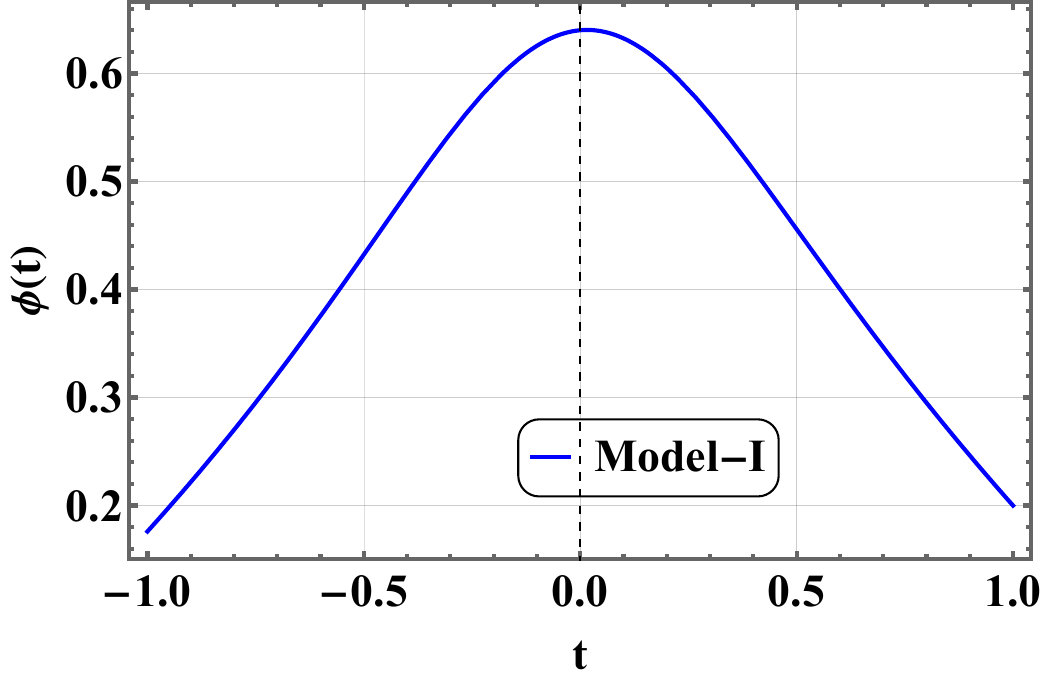} \hfill
    \includegraphics[width=0.45\textwidth]{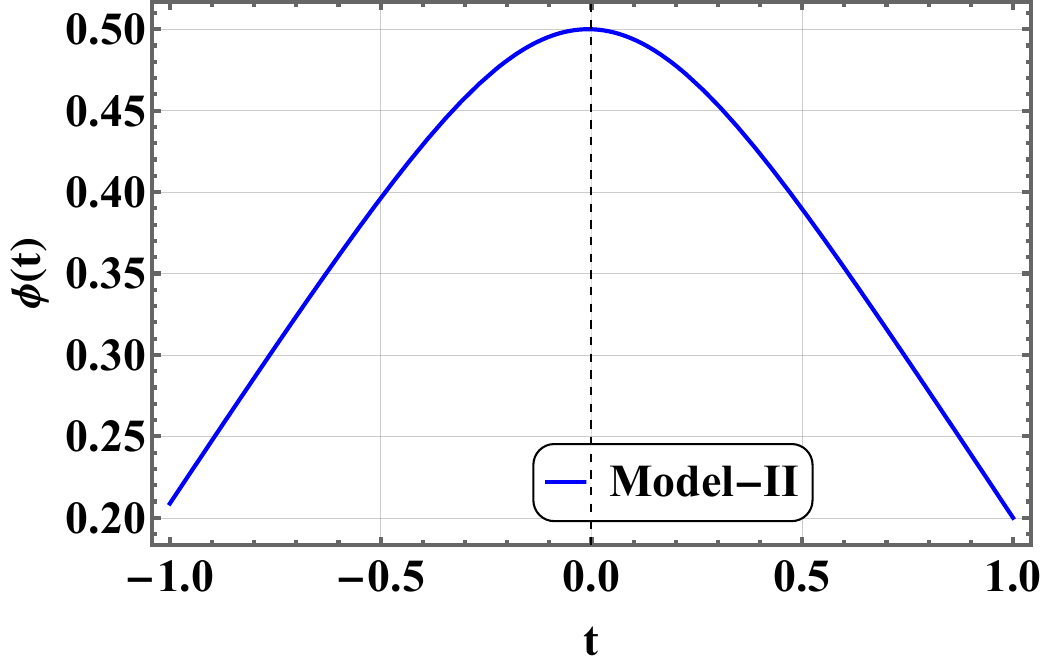}
    \caption{Comparison of scalar field $\phi(t)$ evolution for Model-I and Model-II.}
    \label{fig:phi:models}
\end{figure}

\section{Energy Conditions}\label{ec}
\begin{figure}[H]
    \centering
    \includegraphics[width=0.45\textwidth]{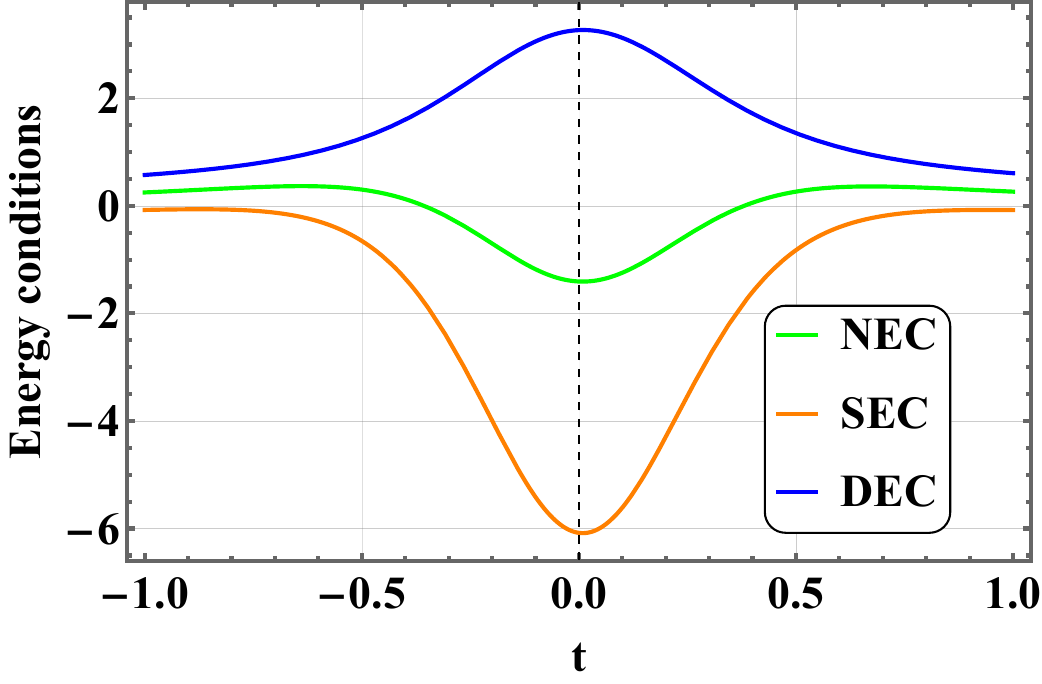} \hfill
    \includegraphics[width=0.45\textwidth]{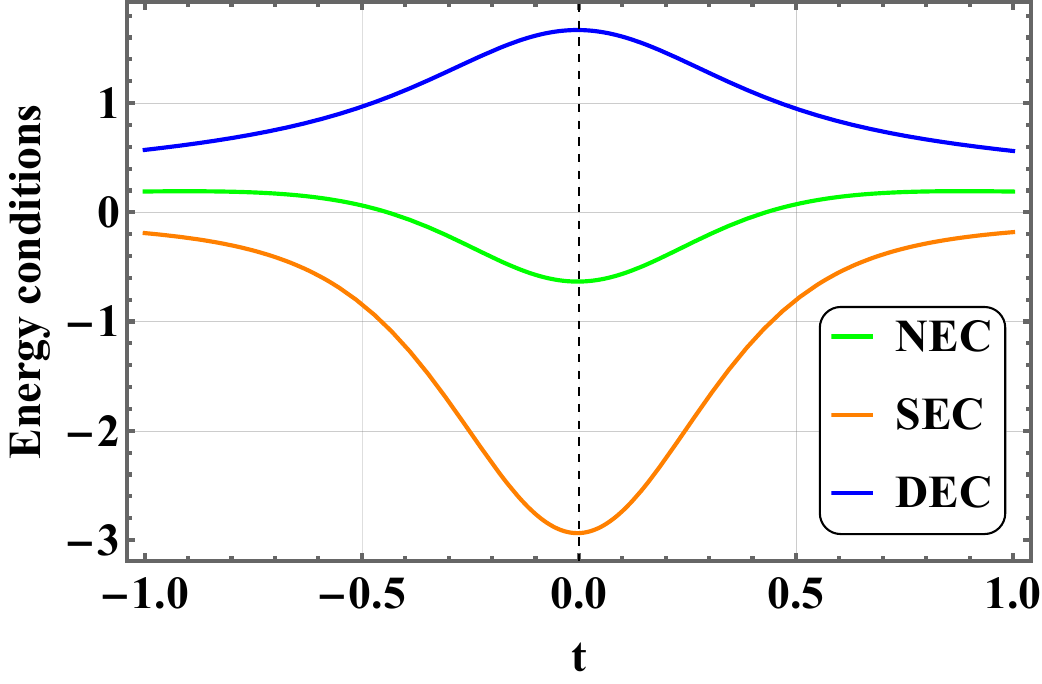}
    \caption{Evolution of energy conditions: Dominant (DEC), Strong (SEC), and Null (NEC) as functions of cosmic time \( t \) for two different models.}
    \label{fig:energy-conditions}
\end{figure}

Energy conditions impose coordinate-invariant constraints on the energy-momentum tensor. The main conditions are \cite{Raychaudhuri-1955, Carroll-2019, Curiel2017}:  
\begin{itemize}
    \item \textbf{Weak Energy Condition (WEC):}  
    $T_{ij}t^{i}t^{j}\geq 0$ for any timelike $t^{i}$.  
    For a perfect fluid:  
    \begin{equation}
        T_{ij}u^{i}u^{j}=\rho, \quad 
        T_{ij}\xi^{i}\xi^{j}=(\rho+p)(u_{i}\xi^{i})^{2}.
    \end{equation}
    Implies $\rho \geq 0$, $\rho+p \geq 0$.  

    \item \textbf{Null Energy Condition (NEC):}  
    $T_{ij}\xi^{i}\xi^{j}\geq 0$ for any null $\xi^{i}$.  
    Equivalent to $\rho+p \geq 0$.  

    \item \textbf{Strong Energy Condition (SEC):}  
    $T_{ij}t^{i}t^{j}-\tfrac{1}{2}T^{k}_{~k}t^{l}t_{l}\geq 0$.  
    Equivalent to $\rho+p \geq 0$, $\rho+3p \geq 0$.  
    Implies attractive gravity.  

    \item \textbf{Dominant Energy Condition (DEC):}  
    $T_{ij}t^{i}t^{j}\geq 0$ and $T^{ij}t_{i}$ is non-spacelike.  
    For a perfect fluid: $\rho \geq |p|$.  
\end{itemize}
To ensure the stability of the model, it is essential to examine the validity of the standard energy conditions. It is observed from Fig. \ref{fig:energy-conditions} that the NEC and the SEC are violated in the vicinity of the bounce epoch, while the DEC remains satisfied throughout the entire cosmic evolution. Furthermore, away from the bounce epoch, the NEC is satisfied, but the SEC continues to be violated at all stages of evolution. Such behavior has been widely reported in various modified gravity frameworks, and its occurrence is consistent with the physical requirements for realizing a non-singular bouncing cosmological scenario.

\section{Conclusion}\label{conclusion}
The bouncing models play a vital role in solving the singularity problem and are analyzed in the modified GR formalism. However, such models have not been explored much in the TG formalism, especially in the presence of a scalar field. This study addresses two well-motivated coupling models $Q=\beta H \dot{\phi}^2$ \cite{Roy_2019, Roy:2023uhc, Ashmita_2024} and $Q=\tau H \rho_{\phi} \dot{\phi}^2$ \cite{Roy:2023uhc, Kumar_2019}, where $Q$ represents the non-homogeneous part of the Klein–Gordon equation in interacting TG. Effective tools, such as energy conditions, are employed to check the viability of these models. The violation of the NEC at the bounce epoch supports the analysis in both cases. Furthermore, the violation of the strong energy condition supports the accelerating behavior of the EoS parameters (see Figs.~\ref{fig:eos:m1}, \ref{fig:eos:m2}). An increase in the value of the coupling coefficient enhances the depth of the EoS curve at the bounce epoch and indicates a phantom region more promptly. The EoS parameter for the second model also show the phantom behaviour at the bounce epoch. We analyse the behaiour for the different parametric values $\tau= 0.6, 0.4, 0.2$, this change in the values are not much affecting the EoS parameter at the bounce epoch. The energy conditions indicate a higher presence at the bounce epoch, supporting the emergence of the bouncing solution at that point Ref. Figs.(\ref{fig:density:m1},\ref{fig:density:m2}). The scalar field traced in both the models Ref. Fig. \ref{fig:phi:models} lie in the positive region as expected, and are the numerical solutions to the Klein-Gordon equation presented in Eq. \eqref{klein_Gordon}. We have also analyzed the co-moving Hubble radius and observed that this supports the choice of the matter bounce scale factor to study bouncing behaviour. Overall, this study establishes two well-defined coupling models to analyze bouncing solutions in interacting TG with a scalar field. Additionally, it motivates the exploration of models that can consistently explain the bouncing nature of cosmic
evolution while bypassing the singularity problem.

\bibliographystyle{utphys}
\bibliography{references}

\providecommand{\href}[2]{#2}\begingroup\raggedright\begin{thebibliography}{10}

\bibitem{weinberg2008}
S.~Weinberg, {\em Cosmology}.
\newblock Oxford University Press, U. K., 2008.

\bibitem{Carroll1992}
S.~M. Carroll, W.~H. Press, and E.~L. Turner, ``The cosmological constant,''
  \href{http://dx.doi.org/10.1146/annurev.aa.30.090192.002435}{{\em Annual
  Review of Astronomy and Astrophysics} {\bf 30} (1992)  499--542},
  \href{http://arxiv.org/abs/astro-ph/0004075}{{\tt arXiv:astro-ph/0004075
  [astro-ph]}}.

\bibitem{A_Guth}
A.~H. Guth, ``Inflationary universe: A possible solution to the horizon and
  flatness problems,'' \href{http://dx.doi.org/10.1103/PhysRevD.23.347}{{\em
  Phys. Rev. D} {\bf 23} (1981)  }.

\bibitem{LINDE1982389}
A.~Linde, ``A new inflationary universe scenario: A possible solution of the
  horizon, flatness, homogeneity, isotropy and primordial monopole problems,''
  \href{http://dx.doi.org/https://doi.org/10.1016/0370-2693(82)91219-9}{{\em
  Phys. Lett. B} {\bf 108} (1982)  }.

\bibitem{Borde_1994}
A.~Borde and A.~Vilenkin, ``Eternal inflation and the initial singularity,''
  \href{http://dx.doi.org/10.1103/physrevlett.72.3305}{{\em Phys. Rev. Lett.}
  {\bf 72} (1994)  }, \href{http://arxiv.org/abs/gr-qc/9312022v1}{{\tt
  arXiv:gr-qc/9312022v1 [gr-qc]}}.

\bibitem{Ashtekar_2006}
A.~Ashtekar, T.~Pawlowski, and P.~Singh, ``Quantum nature of the big bang,''
  \href{http://dx.doi.org/10.1103/physrevlett.96.141301}{{\em Phys. Rev. Lett.}
  {\bf 96} (2006) no.~14, }, \href{http://arxiv.org/abs/gr-qc/0602086}{{\tt
  arXiv:gr-qc/0602086 [gr-qc]}}.

\bibitem{Ilyas_2021}
M.~Ilyas and W.~U. Rahman, ``{Bounce cosmology in $f(R)$ gravity},''
  \href{http://dx.doi.org/10.1140/epjc/s10052-021-08955-7}{{\em Eur. Phys. J.
  C} {\bf 81} (2021) no.~2, }, \href{http://arxiv.org/abs/2102.03612v1}{{\tt
  arXiv:2102.03612v1 [gr-qc]}}.

\bibitem{Peter_2002}
P.~Peter and N.~Pinto-Neto, ``Primordial perturbations in a nonsingular
  bouncing universe model,''
  \href{http://dx.doi.org/10.1103/physrevd.66.063509}{{\em Phys. Rev. D} {\bf
  66} (2002) no.~6, }, \href{http://arxiv.org/abs/hep-th/0203013}{{\tt
  arXiv:hep-th/0203013 [hep-th]}}.

\bibitem{AGRAWAL2021100863}
A.~Agrawal, L.~Pati, S.~Tripathy, and B.~Mishra, ``{Matter bounce scenario and
  the dynamical aspects in $f(Q,T)$ gravity},''
  \href{http://dx.doi.org/10.1016/j.dark.2021.100863}{{\em Phys. Dark Univ.}
  {\bf 33} (2021)  100863}, \href{http://arxiv.org/abs/2108.02575}{{\tt
  arXiv:2108.02575 [gr-qc]}}.

\bibitem{agrawal2022bouncing}
A.~Agrawal, F.~Tello-Ortiz, B.~Mishra, and S.~Tripathy, ``Bouncing cosmology in
  extended gravity and its reconstruction as a dark energy model,''
  \href{http://dx.doi.org/10.1002/prop.202100065}{{\em Fortschritte der Physik}
  {\bf 70} (2022) no.~1, 2100065}, \href{http://arxiv.org/abs/2111.02894}{{\tt
  arXiv:2111.02894 [gr-qc]}}.

\bibitem{agrawal2022role}
A.~Agrawal, S.~Tripathy, S.~Pal, and B.~Mishra, ``Role of extended gravity
  theory in matter bounce dynamics,''
  \href{http://dx.doi.org/10.1088/1402-4896/ac49b2}{{\em Physica Scripta} {\bf
  97} (2022) no.~2, 025002}, \href{http://arxiv.org/abs/2201.03783}{{\tt
  arXiv:2201.03783 [gr-qc]}}.

\bibitem{agrawal2023bouncing}
A.~Agrawal, S.~Mishra, S.~Tripathy, and B.~Mishra, ``{Bouncing Cosmological
  Models in a Functional form of $F(R)$ Gravity},''
  \href{http://dx.doi.org/10.1134/S0202289323030027}{{\em Gravitation and
  Cosmology} {\bf 29} (2023) no.~3, 294--304},
  \href{http://arxiv.org/abs/2210.09726}{{\tt arXiv:2210.09726 [gr-qc]}}.

\bibitem{agrawal2023matter}
A.~Agrawal, B.~Mishra, and P.~Agrawal, ``Matter bounce scenario in extended
  symmetric teleparallel gravity,''
  \href{http://dx.doi.org/10.1140/epjc/s10052-023-11266-8}{{\em Eur. Phys. J.
  C} {\bf 83} (2023) no.~2, 113}, \href{http://arxiv.org/abs/2206.02783}{{\tt
  arXiv:2206.02783 [gr-qc]}}.

\bibitem{Cai_2011}
Y.-F. Cai, S.-H. Chen, J.~B. Dent, S.~Dutta, and E.~N. Saridakis, ``{Matter
  bounce cosmology with the $f(T)$ gravity},''
  \href{http://dx.doi.org/10.1088/0264-9381/28/21/215011}{{\em Class. Quan.
  Grav.} {\bf 28} (2011)  }, \href{http://arxiv.org/abs/1104.434}{{\tt
  arXiv:1104.434 [astro-ph]}}.

\bibitem{Bamba_2014}
K.~Bamba, A.~N. Makarenko, A.~N. Myagky, and S.~D. Odintsov, ``Bouncing
  cosmology in modified gauss–bonnet gravity,''
  \href{http://dx.doi.org/10.1016/j.physletb.2014.04.004}{{\em Phys. Lett. B}
  {\bf 732} (2014)  }, \href{http://arxiv.org/abs/1403.3242}{{\tt
  arXiv:1403.3242 [hep-th]}}.

\bibitem{Amani_2016}
A.~R. Amani, ``{The bouncing cosmology with $f(R)$ gravity and its
  reconstructing},'' \href{http://dx.doi.org/10.1142/s0218271816500711}{{\em
  Int. J. Mod. Phys. D} {\bf 25} (2016) no.~06, },
  \href{http://arxiv.org/abs/1512.03475}{{\tt arXiv:1512.03475 [gr-qc]}}.

\bibitem{dela_2018}
A.~de~la Cruz-Dombriz, G.~Farrugia, J.~L. Said, and
  D.~S\'aez-Chill\'on~G\'omez, ``{Cosmological bouncing solutions in extended
  teleparallel gravity theories},''
  \href{http://dx.doi.org/10.1103/PhysRevD.97.104040}{{\em Phys. Rev. D} {\bf
  97} (2018) no.~10, 104040}, \href{http://arxiv.org/abs/1801.10085}{{\tt
  arXiv:1801.10085 [gr-qc]}}.

\bibitem{caruana2020}
M.~Caruana, G.~Farrugia, and J.~Levi~Said, ``{Cosmological bouncing solutions
  in $f (T, B)$ gravity},''
  \href{http://dx.doi.org/https://doi.org/10.1140/epjc/s10052-020-8204-3}{{\em
  Euro. Phys. J. C} {\bf 80} (2020) no.~7, 640},
  \href{http://arxiv.org/abs/2007.09925}{{\tt arXiv:2007.09925 [gr-qc]}}.

\bibitem{MISHRA2024138968}
B.~Mishra, S.~A. Kadam, and S.~K. Tripathy, ``Scalar field induced dynamical
  evolution in teleparallel gravity,''
  \href{http://dx.doi.org/10.1016/j.physletb.2024.138968}{{\em Phys. Lett. B}
  {\bf 857} (2024)  138968}, \href{http://arxiv.org/abs/2406.15896}{{\tt
  arXiv:2406.15896 [gr-qc]}}.

\bibitem{2022EPJP}
M.~G. {Ganiou}, M.~J.~S. {Houndjo}, C.~{A{\"\i}namon}, L.~{Ayivi}, and
  A.~{Kanfon}, ``{Reconstruction method applied to bounce cosmology and
  inflationary scenarios in cosmological $f(G)$ gravity},''
  \href{http://dx.doi.org/10.1140/epjp/s13360-021-02140-1}{{\em Eur. Phys. J.
  P.} {\bf 137} (2022) no.~2, }.

\bibitem{Gadbail_2023}
G.~N. Gadbail, A.~Kolhatkar, S.~Mandal, and P.~K. Sahoo, ``{Correction to
  Lagrangian for bouncing cosmologies in $f(Q)$ gravity},''
  \href{http://dx.doi.org/10.1140/epjc/s10052-023-11798-z}{{\em Eur. Phys. J.
  C} {\bf 83} (2023)  }, \href{http://arxiv.org/abs/2304.10245}{{\tt
  arXiv:2304.10245 [gr-qc]}}.

\bibitem{Azhar_25}
N.~Azhar, ``{Viability of reconstructed bouncing cosmologies in $f(T,\tau)$
  gravity},'' \href{http://dx.doi.org/10.1142/S0219887824503043}{{\em Inter. J.
  Geom. Meth. Mod. Phys.} {\bf 22} (2025) no.~04, 2450304}.

\bibitem{Maluf:2013gaa}
J.~W. Maluf, ``{The teleparallel equivalent of general relativity},''
  \href{http://dx.doi.org/10.1002/andp.201200272}{{\em Annalen Phys.} {\bf 525}
  (2013)  339--357}, \href{http://arxiv.org/abs/1303.3897}{{\tt arXiv:1303.3897
  [gr-qc]}}.

\bibitem{Pereira:2013qza}
J.~G. Pereira, {\em {Teleparallelism: A New Insight Into Gravity}},
  \href{http://dx.doi.org/10.1007/978-3-642-41992-8_11}{pp.~197--212}.
\newblock Springer Berlin Heidelberg, 2014.
\newblock \href{http://arxiv.org/abs/1302.6983}{{\tt arXiv:1302.6983 [gr-qc]}}.

\bibitem{Ferraro:2008ey}
R.~Ferraro and F.~Fiorini, ``{On Born-Infeld Gravity in Weitzenbock
  spacetime},'' \href{http://dx.doi.org/10.1103/PhysRevD.78.124019}{{\em Phys.
  Rev. D} {\bf 78} (2008)  124019}, \href{http://arxiv.org/abs/0812.1981}{{\tt
  arXiv:0812.1981 [gr-qc]}}.

\bibitem{Weitzenbock1923}
R.~Weitzenb\"{o}ock, {\em `Invariantentheorie'}.
\newblock Noordhoff, Gronningen, 1923.

\bibitem{Aldrovandi:2013wha}
R.~Aldrovandi and J.~G. Pereira,
  \href{http://dx.doi.org/10.1007/978-94-007-5143-9}{{\em {Teleparallel
  Gravity}: {An Introduction}}}.
\newblock Springer, 2013.

\bibitem{bahamonde:2021teleparallel}
S.~Bahamonde {\em et al.}, ``Teleparallel gravity: from theory to cosmology,''
  \href{http://dx.doi.org/https://doi.org/10.1088/1361-6633/ac9cef}{{\em Rep.
  Prog. Phys.} {\bf 86} (2023)  207pp},
  \href{http://arxiv.org/abs/2106.13793}{{\tt arXiv:2106.13793 [gr-qc]}}.

\bibitem{Di_Valentino_2025}
E.~Di~Valentino {\em et al.}, ``The cosmoverse white paper: Addressing
  observational tensions in cosmology with systematics and fundamental
  physics,'' \href{http://dx.doi.org/10.1016/j.dark.2025.101965}{{\em Phys.
  Dark Univ.} {\bf 49} (2025)  }, \href{http://arxiv.org/abs/2504.01669}{{\tt
  arXiv:2504.01669 [astro-ph]}}.

\bibitem{2011JCAP}
Y.~{Zhang}, H.~{Li}, Y.~{Gong}, and Z.-H. {Zhu}, ``{Notes on $f(T)$
  theories},'' \href{http://dx.doi.org/10.1088/1475-7516/2011/07/015}{{\em
  \jcap} {\bf 2011} (2011) no.~7, 015},
  \href{http://arxiv.org/abs/1103.0719}{{\tt arXiv:1103.0719 [astro-ph.CO]}}.

\bibitem{DUCHANIYA2024f(T)}
{L. K. Duchaniya and Kanika Gandhi and B. Mishra}, ``{Attractor behavior of
  $f(T)$ modified gravity and the cosmic acceleration},''
  \href{http://dx.doi.org/https://doi.org/10.1016/j.dark.2024.101461}{{\em
  Phys. Dark Univ.} {\bf 44} (2024)  101461},
  \href{http://arxiv.org/abs/2303.09076}{{\tt arXiv:2303.09076 [gr-qc]}}.

\bibitem{Bahamonde:2015zma}
S.~Bahamonde, C.~G. B\"ohmer, and M.~Wright, ``{Modified teleparallel theories
  of gravity},'' \href{http://dx.doi.org/10.1103/PhysRevD.92.104042}{{\em Phys.
  Rev. D} {\bf 92} (2015) no.~10, 104042},
  \href{http://arxiv.org/abs/1508.05120}{{\tt arXiv:1508.05120 [gr-qc]}}.

\bibitem{Wang_2020}
D.~Wang and D.~Mota, ``{Can $f(T)$ gravity resolve $H_{0}$ the tension?},''
  \href{http://dx.doi.org/10.1103/physrevd.102.063530}{{\em Phys. Rev. D} {\bf
  102} (2020)  }, \href{http://arxiv.org/abs/2003.10095}{{\tt arXiv:2003.10095
  [astro-ph]}}.

\bibitem{Emmanuel_H}
E.~N. Saridakis, ``{Solving both $H_{0}$ and $\sigma_{8}$ tensions in $f(T)$
  gravity},'' \href{http://dx.doi.org/10.1142/9789811269776_0139}{{\em The
  Sixteenth Marcel Grossmann Meeting} (2023)  1783--1791},
  \href{http://arxiv.org/abs/2301.06881}{{\tt arXiv:2301.06881 [gr-qc]}}.

\bibitem{Bengochea:2008gz}
G.~R. Bengochea and R.~Ferraro, ``{Dark torsion as the cosmic speed-up},''
  \href{http://dx.doi.org/10.1103/PhysRevD.79.124019}{{\em Phys. Rev. D} {\bf
  79} (2009)  124019}, \href{http://arxiv.org/abs/0812.1205}{{\tt
  arXiv:0812.1205 [astro-ph]}}.

\bibitem{Iorio:2012cm}
L.~Iorio and E.~N. Saridakis, ``{Solar system constraints on $f(T)$ gravity},''
  \href{http://dx.doi.org/10.1111/j.1365-2966.2012.21995.x}{{\em Mon. Not. Roy.
  Astron. Soc.} {\bf 427} (2012)  1555},
  \href{http://arxiv.org/abs/1203.5781}{{\tt arXiv:1203.5781 [gr-qc]}}.

\bibitem{Farrugia:2016xcw}
G.~Farrugia, J.~Levi~Said, and M.~L. Ruggiero, ``{Solar System tests in $f(T)$
  gravity},'' \href{http://dx.doi.org/10.1103/PhysRevD.93.104034}{{\em Phys.
  Rev. D} {\bf 93} (2016) no.~10, 104034},
  \href{http://arxiv.org/abs/1605.07614}{{\tt arXiv:1605.07614 [gr-qc]}}.

\bibitem{Miao_2011}
R.-X. Miao, M.~Li, and Y.-G. Miao, ``{Violation of the first law of black hole
  thermodynamics in $f(T)$ gravity},''
  \href{http://dx.doi.org/10.1088/1475-7516/2011/11/033}{{\em JCAP} {\bf 2011}
  (2011) no.~11, }, \href{http://arxiv.org/abs/1107.0515}{{\tt arXiv:1107.0515
  [hep-th]}}.

\bibitem{Basilakos:2013rua}
S.~Basilakos, S.~Capozziello, M.~De~Laurentis, A.~Paliathanasis, and
  M.~Tsamparlis, ``{Noether symmetries and analytical solutions in
  $f(T)$-cosmology: A complete study},''
  \href{http://dx.doi.org/10.1103/PhysRevD.88.103526}{{\em Phys. Rev. D} {\bf
  88} (2013)  103526}, \href{http://arxiv.org/abs/1311.2173}{{\tt
  arXiv:1311.2173 [gr-qc]}}.

\bibitem{ODINTSOV_2015}
S.~Odintsov, V.~Oikonomou, and E.~N. Saridakis, ``{Superbounce and loop quantum
  ekpyrotic cosmologies from modified gravity: $F(R)$, $F(G)$ and $F(T)$
  theories},''
  \href{http://dx.doi.org/https://doi.org/10.1016/j.aop.2015.08.021}{{\em
  Annals of Physics} {\bf 363} (2015)  141--163},
  \href{http://arxiv.org/abs/1501.06591}{{\tt arXiv:1501.06591 [gr-qc]}}.

\bibitem{Copeland_1998}
E.~J. Copeland, A.~R. Liddle, and D.~Wands, ``Exponential potentials and
  cosmological scaling solutions,''
  \href{http://dx.doi.org/10.1103/physrevd.57.4686}{{\em Phys. Rev. D} {\bf 57}
  (1998)  }, \href{http://arxiv.org/abs/gr-qc/9711068}{{\tt arXiv:gr-qc/9711068
  [gr-qc]}}.

\bibitem{Ashmita_2024}
Ashmita, K.~Banerjee, and P.~K. Das, ``Constructing viable interacting dark
  matter and dark energy models: a dynamical systems approach,''
  \href{http://dx.doi.org/10.1088/1475-7516/2024/11/034}{{\em JCAP} {\bf 2024}
  (2024)  }, \href{http://arxiv.org/abs/2410.02261}{{\tt arXiv:2410.02261
  [gr-qc]}}.

\bibitem{Roy:2023uhc}
N.~Roy, ``{Exploring the possibility of interacting quintessence model as an
  alternative to the $\Lambda $CDM model},''
  \href{http://dx.doi.org/10.1007/s10714-023-03160-1}{{\em Gen. Rel. Grav.}
  {\bf 55} (2023) no.~10, 115}, \href{http://arxiv.org/abs/2302.10509}{{\tt
  arXiv:2302.10509 [astro-ph.CO]}}.

\bibitem{G.Otalora2013JCAPDSA}
G.~Otalora, ``Scaling attractors in interacting teleparallel dark energy,''
  \href{http://dx.doi.org/10.1088/1475-7516/2013/07/044}{{\em JCAP} {\bf 2013}
  (2013)  }, \href{http://arxiv.org/abs/1305.0474}{{\tt arXiv:1305.0474
  [gr-qc]}}.

\bibitem{WEI2012430}
H.~Wei, ``Dynamics of teleparallel dark energy,''
  \href{http://dx.doi.org/10.1016/j.physletb.2012.05.006}{{\em Phys. Lett. B}
  {\bf 712} (2012)  }, \href{http://arxiv.org/abs/1109.6107}{{\tt
  arXiv:1109.6107 [gr-qc]}}.

\bibitem{Xu_2012}
C.~Xu, E.~N. Saridakis, and G.~Leon, ``Phase-space analysis of teleparallel
  dark energy,'' \href{http://dx.doi.org/10.1088/1475-7516/2012/01/002}{{\em
  JCAP} {\bf 2012} (2012)  }, \href{http://arxiv.org/abs/1202.3781}{{\tt
  arXiv:1202.3781 [gr-qc]}}.

\bibitem{Geng:2011}
C.-Q. Geng, C.-C. Lee, E.~N. Saridakis, and Y.-P. Wu, ``Teleparallel dark
  energy,'' \href{http://dx.doi.org/10.1016/j.physletb.2011.09.082}{{\em Phys.
  Lett. B} {\bf 704} (2011)  384–387},
  \href{http://arxiv.org/abs/1109.1092v2}{{\tt arXiv:1109.1092v2 [hep-th]}}.

\bibitem{Farrugia:2018gyz}
G.~Farrugia, J.~Levi~Said, V.~Gakis, and E.~N. Saridakis, ``Gravitational waves
  in modified teleparallel theories,''
  \href{http://dx.doi.org/10.1103/PhysRevD.97.124064}{{\em Phys. Rev. D} {\bf
  97} (2018) no.~12, 124064}, \href{http://arxiv.org/abs/1804.07365}{{\tt
  arXiv:1804.07365 [gr-qc]}}.

\bibitem{Krssak:2015oua}
M.~Kr\v{s}\v{s}\'ak and E.~N. Saridakis, ``{The covariant formulation of $f(T)$
  gravity},'' \href{http://dx.doi.org/10.1088/0264-9381/33/11/115009}{{\em
  Class. Quant. Grav.} {\bf 33} (2016) no.~11, 115009},
  \href{http://arxiv.org/abs/1510.08432}{{\tt arXiv:1510.08432 [gr-qc]}}.

\bibitem{Bahamonde:2021gfp}
S.~Bahamonde {\em et al.}, ``Teleparallel gravity: from theory to cosmology,''
  \href{http://dx.doi.org/https://doi.org/10.1088/1361-6633/ac9cef}{{\em Rep.
  Prog. Phys.} {\bf 86} (2023)  }, \href{http://arxiv.org/abs/2106.13793}{{\tt
  arXiv:2106.13793 [gr-qc]}}.

\bibitem{roy2018dynamical}
N.~Roy and N.~Bhadra, ``Dynamical systems analysis of phantom dark energy
  models,'' \href{http://dx.doi.org/10.1088/1475-7516/2018/06/002}{{\em JCAP}
  {\bf 2018} (2018)  002}, \href{http://arxiv.org/abs/1710.05968}{{\tt
  arXiv:1710.05968 [gr-qc]}}.

\bibitem{Gumjudpai_2005}
B.~Gumjudpai, T.~Naskar, M.~Sami, and S.~Tsujikawa, ``Coupled dark energy:
  towards a general description of the dynamics,''
  \href{http://dx.doi.org/10.1088/1475-7516/2005/06/007}{{\em JCAP} {\bf 2005}
  (2005) no.~06, }, \href{http://arxiv.org/abs/hep-th/0502191}{{\tt
  arXiv:hep-th/0502191 [hep-th]}}.

\bibitem{Lohakare_2022}
S.~Lohakare, F.~Tello-Ortiz, S.~Tripathy, and B.~Mishra, ``Bouncing cosmology
  in modified gravity with higher-order gauss–bonnet curvature term,''
  \href{http://dx.doi.org/10.3390/universe8120636}{{\em Universe} {\bf 8}
  (2022) no.~12, }, \href{http://arxiv.org/abs/2211.03609}{{\tt
  arXiv:2211.03609 [gr-qc]}}.

\bibitem{Haro_2012}
J.~d. Haro, ``Future singularity avoidance in phantom dark energy models,''
  \href{http://dx.doi.org/10.1088/1475-7516/2012/07/007}{{\em JCAP} {\bf 2012}
  (2012) no.~07, }, \href{http://arxiv.org/abs/1204.5604}{{\tt arXiv:1204.5604
  [gr-qc]}}.

\bibitem{de_Haro_2015}
J.~de~Haro and J.~Amorós, ``Viability of the matter bounce scenario,'' {\em J
  Phys.: Conf. Ser.} {\bf 600} (2015)  ,
  \href{http://arxiv.org/abs/1411.7611}{{\tt arXiv:1411.7611 [gr-qc]}}.

\bibitem{Odintsov-2020-959}
S.~Odintsov, V.~Oikonomou, and T.~Paul, ``{Bottom-up reconstruction of
  non-singular bounce in $F(R)$ gravity from observational indices},''
  \href{http://dx.doi.org/10.1016/j.nuclphysb.2020.115159}{{\em Nucl. Phys. B}
  {\bf 959} (2020)  115159}, \href{http://arxiv.org/abs/2008.13201}{{\tt
  arXiv:2008.13201 [gr-qc]}}.

\bibitem{Chen_2009}
X.-m. Chen and Y.~Gong, ``Fixed points in interacting dark energy models,''
  \href{http://dx.doi.org/10.1016/j.physletb.2009.03.064}{{\em Phys. Lett. B}
  {\bf 675} (2009) no.~1, }, \href{http://arxiv.org/abs/0811.1698}{{\tt
  arXiv:0811.1698 [gr-qc]}}.

\bibitem{B_hmer_2008}
C.~G. Böhmer, G.~Caldera-Cabral, R.~Lazkoz, and R.~Maartens, ``Dynamics of
  dark energy with a coupling to dark matter,''
  \href{http://dx.doi.org/10.1103/physrevd.78.023505}{{\em Phys. Rev. D} {\bf
  78} (2008) no.~2, }, \href{http://arxiv.org/abs/0801.1565}{{\tt
  arXiv:0801.1565 [gr-qc]}}.

\bibitem{Mimoso_2006}
J.~P. Mimoso, A.~Nunes, and D.~Pavón, ``Asymptotic behavior of the warm
  inflation scenario with viscous pressure,''
  \href{http://dx.doi.org/10.1103/physrevd.73.023502}{{\em Phys. Rev. D} {\bf
  73} (2006) no.~2, }, \href{http://arxiv.org/abs/gr-qc/0512057}{{\tt
  arXiv:gr-qc/0512057 [gr-qc]}}.

\bibitem{Roy_2019}
N.~Roy and K.~Bamba, ``Arbitrariness of potentials in interacting quintessence
  models,'' \href{http://dx.doi.org/10.1103/physrevd.99.123520}{{\em Phys. Rev.
  D} {\bf 99} (2019) no.~12, }, \href{http://arxiv.org/abs/1811.03234}{{\tt
  arXiv:1811.03234 [astro-ph]}}.

\bibitem{Kumar_2019}
S.~Kumar, R.~C. Nunes, and S.~K. Yadav, ``Dark sector interaction: a remedy of
  the tensions between cmb and lss data,''
  \href{http://dx.doi.org/10.1140/epjc/s10052-019-7087-7}{{\em Europ. Phys. J.
  C} {\bf 79} (2019)  }, \href{http://arxiv.org/abs/1903.04865}{{\tt
  arXiv:1903.04865 [astro-ph]}}.

\bibitem{Raychaudhuri-1955}
A.~Raychaudhuri, ``{Relativistic Cosmology. I},''
  \href{http://dx.doi.org/10.1103/PhysRev.98.1123}{{\em Phys. Rev.} {\bf 98}
  (1955)  1123--1126}.

\bibitem{Carroll-2019}
S.~M. Carroll, \href{http://dx.doi.org/10.1017/9781108770385}{{\em Spacetime
  and geometry: An introduction to general relativity}}.
\newblock Cambridge University Press, 2019.

\bibitem{Curiel2017}
E.~Curiel, \href{http://dx.doi.org/10.1007/978-1-4939-3210-8_3}{{\em A primer
  on energy conditions}}.
\newblock Springer New York, New York, 2017.
\newblock \href{http://arxiv.org/abs/1405.0403}{{\tt arXiv:1405.0403
  [physics]}}.

\end{thebibliography}\endgroup
\end{document}